\documentclass[aps,twocolumn,floats,superscriptaddress,prd,nofootinbib]{revtex4-1}
\usepackage{graphicx, epsfig, bm, amsmath}
\def\mnras{Mon.~Not.~Roy.~Astron.~Soc.}
\usepackage{color}
\usepackage{hyperref}
\begin{document}


\newcommand{\sfig}[2]{
\includegraphics[width=#2]{#1}
                }
\newcommand{\Sfig}[2]{
        \begin{figure}[thbp]
        \sfig{#1.eps}{\columnwidth}
        \caption{{\small #2}}
        \label{fig:#1}
        \end{figure}
}
\newcommand{\Rf}[1]{\ref{fig:#1}}
\newcommand{\rf}[1]{\ref{fig:#1}}
\def\be{\begin{equation}}
\def\ee{\end{equation}}
\def\bea{\begin{eqnarray}}
\def\eea{\end{eqnarray}}
\newcommand{\vs}{\nonumber\\}
\newcommand{\ec}[1]{Eq.~(\ref{eqn:#1})}
\newcommand{\eec}[2]{Eqs.~(\ref{eqn:#1}) and (\ref{eqn:#2})}
\newcommand{\Ec}[1]{(\ref{eqn:#1})}
\newcommand{\eql}[1]{\label{eqn:#1}}

\newcommand{\lcdm}{$\Lambda$CDM}

\newcommand\fcoll{f_{\rm coll}}

\newcommand{\fnl}{f_{\rm NL}}

\def\bdm{\begin{displaymath}}
\def\edm{\end{displaymath}}
\def\curv{\mathcal{R}}
\def\hdelta{\delta}
\def\colcom{f_{\rm c}}
\def\({\left(}
\def\){\right)}
\def\[{\left[}
\def\]{\right]}

\definecolor{darkgreen}{cmyk}{0.85,0.2,1.00,0.2}
\definecolor{darkorange}{cmyk}{0.0,0.47,1.00,0.0}
\newcommand{\peter}[1]{\textcolor{red}{[{\bf PA}: #1]}}
\newcommand{\scott}[1]{\textcolor{blue}{[{\bf SD}: #1]}}
\newcommand{\adam}[1]{\textcolor{darkgreen}{[{\bf AL}: #1]}}
\newcommand{\eric}[1]{\textcolor{darkorange}{[{\bf EB}: #1]}}

\newcommand{\aap}{Astron. Astrophys.}


\pagestyle{plain}

\title{Non-Gaussianity and Excursion Set Theory: Halo Bias}

\author{Peter Adshead}
\affiliation{Kavli Institute for Cosmological Physics,  Enrico Fermi Institute, University of Chicago, Chicago, Illinois 60637}
        
\author{  Eric J Baxter}
\affiliation{Department of Astronomy \& Astrophysics, University of Chicago, Chicago Illinois 60637}

\author{Scott Dodelson}
\affiliation{Kavli Institute for Cosmological Physics,  Enrico Fermi Institute, University of Chicago, Chicago, Illinois 60637}
\affiliation{Department of Astronomy \& Astrophysics, University of Chicago, Chicago Illinois 60637}
\affiliation{Fermilab Center for Particle Astrophysics, Fermi National Accelerator Laboratory, Batavia, Illinois 60510-0500}

\author{Adam Lidz}
\affiliation{Department of Physics \& Astronomy, University of Pennsylvania, 209 South 33rd Street, Philadelphia, Pennsylvania 19104}

\date{\today}

\begin{abstract}
We study the impact of primordial non-Gaussianity generated during inflation on the bias of halos using excursion set theory. We recapture the familiar result that the bias scales as $k^{-2}$ on large  scales for local type non-Gaussianity but explicitly identify the approximations that go into this conclusion and the corrections to it. We solve the more complicated problem of non-spherical halos, for which the collapse threshold is scale dependent.
\end{abstract}

\maketitle

\section{Introduction} 

The standard cosmological model that fits a wide variety of observations is based on inflation, in particular on the production of a nearly scale-invariant spectrum of adiabatic perturbations at very early times. Inflation also predicts, and this too is verified by the data, that the perturbations should be nearly Gaussian. While inflation is successful in explaining the current suite of observations, it has not been successfully connected to the rest of physics. Equivalently, the mechanism that drove inflation has not been identified. 
For these purposes, the small deviations from scale-invariance or Gaussianity may prove crucial~\cite{Komatsu:2009kd}. The simplest single field slow-roll models of inflation predict negligible non-Gaussianity, so a detection has the potential to rule out a large class of models. Multiple-field models, on the other hand, often predict levels of non-Gaussianity within the range of upcoming surveys.

The cosmic microwave background is the most obvious place to search for non-Gaussianity, as the perturbations are observed when they are still small, and therefore relatively unprocessed. Large scale structure, on the other hand, is comprised of highly evolved perturbations that are nonlinear and hence different Fourier modes have mixed with one another. Even if the primordial perturbations were perfectly Gaussian, the observed large scale structure would be highly non-Gaussian. The trick to using large scale structure is to identify a feature in the spectrum that can be caused by primordial non-Gaussianity but not by standard gravitational instability. Recently, such a feature has been identified~\cite{Dalal:2007cu} in the form of scale-dependent bias.\footnote{Generically effects that scale as $k^{-2}$, mimicking the scale dependent bias on large scales resulting from primordial non-Gaussianity, can arise from relativistic effects in general relativity with Gaussian density fluctuations. However, the effects due to primordial non-Gaussianity are much stronger than those arising dynamically from relativistic effects  (see e.g.\ \cite{Bartolo:2010ec, Jeong:2011as, Baldauf:2011bh, Bruni:2011ta}).} While it has been known for some time that primordial non-Gaussianity affects the abundances of rare objects \cite{Matarrese:2000iz,Verde:2000vr, LoVerde:2007ri}, the signal is expected to be diminished due to the gravitational evolution generically evolving the distribution away from Gaussian.

The initial argument for scale-dependent bias was based on both simulations and the statistics of high peak regions~\cite{Press:1973iz,Kaiser:1984sw,Dalal:2007cu,Matarrese:2008nc,Slosar:2008hx,Afshordi:2008ru,McDonald:2008sc,Giannantonio:2009ak}. Subsequently, several groups~\cite{Desjacques:2010gz, Desjacques:2011mq,Scoccimarro:2011pz} have applied the peak-background split in the context of excursion set theory~\cite{Bond:1990iw}. Here we apply a recent generalization of the excursion set approach~\cite{Maggiore:2009rv,Maggiore:2009rx} to the problem of the clustering of halos in order to extract the scale-dependent bias term and to understand under what conditions it holds. We extract the large-scale $k^{-2}$ behavior of the bias for the case of spherical collapse and also generalize to ellipsoidal collapse. Here too the bias contains a scale-dependent $k^{-2}$ piece, but the coefficient differs from that obtained assuming spherical collapse. 

This paper is organized as follows. In \S\ref{review}, we briefly review the path integral approach to the excursion set before we derive the conditional and unconditional crossing rates in \S\ref{sec:uncondcross} and \S\ref{sec:condcross} respectively, and finally, the halo bias in \S\ref{sec:halobias}. In \S\ref{sec:scaledepbias} we extract the bias parameters. We conclude in \S \ref{sec:conclusions}. In Appendix \ref{app:linearbarrier}, we compare our results to a case where the probability density can be evaluated exactly -- where the barrier depends only linearly on time.

\section{Path integral approach to the excursion set}\label{review}

In this section we briefly describe the path integral approach to the excursion set and establish our conventions. The excursion set theory was developed in a seminal paper by Bond, Cole, Efstathiou and Kaiser \cite{Bond:1990iw}. For a nice, accessible review, see Zentner  \cite{Zentner:2006vw}. The path integral approach to the excursion set was developed in a series of papers by Maggiore and Riotto \cite{Maggiore:2009rv,Maggiore:2009rw,Maggiore:2009rx} and extended by De Simone, Maggiore and Riotto to include moving barriers and conditional probabilities \cite{DeSimone:2010mu, DeSimone:2011dn}.

As is usual in excursion set theory, we consider the density fluctuation with respect to the average density $\bar\rho$,
\begin{align}
\delta({\bf x}) = \frac{\rho({\bf x}) - \bar{\rho}}{\bar{\rho}},
\end{align}
smoothed over a region of radius $R$:
\begin{align}
\delta_{R}({\bf x}) = \int d^{3}x' W_{R}({\bf x} - {\bf x}')\delta({\bf x}') = \int \frac{d^{3}k}{(2\pi)^3}\tilde W_{R}(k)\delta_{\bf k}e^{i {\bf k}\cdot {\bf x}}.
\eql{defdel}
\end{align}
In this expression, $ W_{R}({\bf x} - {\bf x}')$ is a window function with characteristic radius $R$, and $\tilde W_R(k)$ is its Fourier transform. Translation invariance implies we can choose ${\bf x}$, and thus for convenience we take ${\bf x} = 0$, and suppress the argument on the smoothed density field from now on, $\delta_{R} \equiv \delta_{R}({\bf x} = 0)$ . Following the notation of \cite{Maggiore:2009rv}, we denote by $S_R = \sigma^{2}(R)$ the variance of the smoothed field $\delta_{R}$,
\begin{align}\
S_R = \int_{0}^{\infty}d\ln k\; \frac{k^{3}}{2\pi^2}\vert \tilde W_R(k)\vert^2 P(k) 
\end{align}
where $P(k)$ is the matter power spectrum.
As shown by Bond et.\ al\ \cite{Bond:1990iw}, if the density fluctuations, $\delta_R$, are purely Gaussian, and the window function is a top-hat in momentum space, $\tilde{W}_{R}(k) = \Theta(k_R - k)$ where $k_R = 1/R$, then the evolution of $\delta_R$ when considered a function of `pseudo-time', $S_R$, is Markovian. That is, each time step is uncorrelated with the previous one, and the future evolution of $\delta_R$ is independent of its history. In this case one can show that the probability density that the field $\delta_R$ takes a value $\delta_{R'}$ when smoothed on a scale $R'$ satisfies the Fokker-Planck equation, with the variance, $S_{R}$, playing the role of pseudo-time (see \cite{Zentner:2006vw} for a review). In order that one does not suffer from the so-called `cloud-in-cloud' problem, an absorbing barrier condition is imposed as a boundary condition on the probability density at the critical threshold for collapse. This means that the probability distribution at time $S_R$ includes only trajectories that never crossed the critical threshold at earlier\footnote{Recall that the variance monotonically increases with decreasing $R$, so ``early'' times correspond to large scales.} times $S_{R'}<S_R$.

Obtaining the probability distribution in the Gaussian-Markovian limit simply amounts to solving the Fokker-Planck equation with various boundary conditions. However, in the general non-Markovian limit one must solve a complicated integro-differential equation for the probability density \cite{Maggiore:2009rv}. Moreover, even in the Gaussian-Markovian limit, it is difficult to obtain solutions for the probability distribution in the case where the absorbing barrier conditions are more complicated than linear in the pseudo-time variable. 

Until recently it was analytically intractable to move far beyond the constant collapse threshold and sharp k-space filtering assumptions of the original formulation of the excursion set \cite{Bond:1990iw}. Using the original techniques, more complicated collapse thresholds have been considered in order to describe collapse scenarios beyond spherical, and  the formation of ionized regions during reionization \cite{Sheth:2001dp, Furlanetto:2004nh}. However, in order to  capture the effects of non-Gaussian fluctuations and more physically realistic filter functions, new techniques for solving for the probability distribution when the variable $\delta_R$ does not evolve in a Markovian fashion with $S_R$ were needed. In recent years there have been several breakthroughs along this avenue \cite{Maggiore:2009rv,Maggiore:2009rw, Maggiore:2009rx,DeSimone:2010mu, DeSimone:2011dn, Paranjape:2011ak, Musso:2012qk, Musso:2012ch}. 

In this work we will follow the path integral approach of Maggiore and Riotto  \cite{Maggiore:2009rv,Maggiore:2009rw}. 
Their formalism is general enough to account for non-Gaussian correlations between the fluctuations \cite{Maggiore:2009rx}, for critical thesholds that evolve with the smoothing scale and for window functions $\tilde{W}_{R}(k)$ other than k-space top hats. Rather than attempting to solve a differential equation for the probability distribution, in the approach of Maggiore and Riotto one directly constructs this quantity by summing over paths that never exceeded the threshold, that is, by performing a path integral. 

Following Maggiore and Riotto, we consider the smoothed density field, $\delta(S) (\equiv \delta_{S_R})$, as a stochastic variable with zero mean $\langle \delta(S) \rangle = 0$. This variable evolves stochastically with time $S$, and we refer to single realization of a $\delta(S)$ as a trajectory. We then consider an ensemble of trajectories of this stochastic variable all starting from the same initial point $\delta(S_{0} = 0) = 0$, and we follow them for some time $S$. We discretize the time interval $[0,S]$ into steps, $\Delta S = \epsilon$, so that $S_k = k\epsilon$, where $k \in \{ 0, 1, .. ,n\}$, and $S_{n} = S$. A discretized trajectory then, is defined as a set of values $\{ \delta_{1}, \delta_{2}, ..., \delta_{n}\}$ so that $\delta(S_{i}) = \delta_{i}$ .

In this path integral approach, the fundamental object is the probability density in this space of trajectories, which is given by
\begin{align}\label{eqn:trajprobdens}
W(\delta_0; \delta_1, ... ,\delta_n ; S_n) \equiv \langle \delta^D(\delta(S_{1})-\delta_1)\ldots\delta^D(\delta(S_{n})-\delta_n)\rangle
\end{align}
where $\delta^D$ is the Dirac delta, which we explicitly denote to avoid confusion with the density perturbation. 
Note that $W$ is defined for all possible values of $\delta$ including those above the threshold.
The below-threshold requirement is enforced at the next stage when computing $\Pi(\delta_n,S_n)$, the probability of arriving at point $\delta_n$ at time $S_n$ through trajectories that have never exceeded some threshold, $B(S_i) \equiv B_i$
\begin{align}\label{eqn:probabilityofdelta}
\Pi(\delta_n,S_n) \equiv \int_{-\infty}^{B_1}\!\!\!\!d\delta_{1}\ldots\int_{-\infty}^{B_{n-1}}\!\!\!\!\!d\delta_{n - 1}W(\delta_0; \delta_1, ... ,\delta_n ; S_n).
\end{align}
The upper limits on each $\delta$ integral allow for the possibility of scale-dependent barriers. Although the threshold value of $\delta$ is constant for spherical collapse (equal to 1.686 today), in non-spherical collapse and other applications, the barrier is generally scale-dependent. 

The fundamental quantities, $W$ and $\Pi$, are the ingredients needed to connect to observations. The fraction of the universe contained in collapsed objects on scales larger than $S_n$ is found from Eq.\ (\ref{eqn:probabilityofdelta}) by integrating over all below-threshold values of $\delta_n$ and taking the complement:
\begin{align}\label{eqn:collapse}
f_{\rm coll}(S_n) = 1- \int_{-\infty}^{B_n}d\delta_n \Pi(\delta_0; \delta_n;S_n).
\end{align}
Differentiating this gives the formation rate of objects collapsing on the scale $S_n$:
\begin{align}
\mathcal{F}(S_n) \equiv  \frac{\partial f_{\rm coll}(S_n) }{\partial S_n}.
\eql{form}
\end{align}
This crossing rate is the {\it unconditional} rate that allows for trajectories passing through any intermediate values of $\delta$. It is also useful to compute a {\it conditional} rate that fixes one of the intermediate values, as $\mathcal{F}(S_n | \delta_m, S_m)$, where the second arguments dictate the time and value of the intermediate fixed point. 
The conditional crossing rate is found from the analogous equation to Eq. (\ref{eqn:collapse}), where the unconditional probability that a trajectory reaches a point $\delta_n$ is replaced by the corresponding conditional probability that a trajectory reached a point $\delta_n$ at time $S_n$ having had a value $\delta_m$ at intermediate time $S_m$ \cite{Ma:2010ep, DeSimone:2011dn},
\begin{align}\label{eqn:probabilityofdeltacond}\nonumber
\Pi &(\delta_n,S_n | \delta_m, S_m)\equiv \\ & \frac{\int_{-\infty}^{B_1}\!\!d\delta_{1}\ldots \widehat{d\delta}_{m}\ldots\int_{-\infty}^{B_{n-1}}\!\!d\delta_{n - 1}W(\delta_0 ; \delta_1, ... ,\delta_n ; S_n)}{\int_{-\infty}^{B_1}\!\!d\delta_{1}\ldots\int_{-\infty}^{B_{m-1}}\!\!d\delta_{m - 1}W(\delta_0 ; \delta_1, ... ,\delta_m ; S_m)}.
\end{align}
The $ \widehat{d\delta}_{m}$ in this expression indicates that we do not integrate over the variable $\delta_m$. 

The ratio of the conditional and unconditional rates quantifies the fact that halos (on scale $S_n$) are more likely to form in overdense regions (on scale $S_m<S_n$) than in an average place in the universe. The halo overdensity on large scales $S_m$ in initial Lagrangian space is \cite{Mo:1995cs}
\begin{equation}
\label{eqn:halooverdensity}
1+\delta_m^{\rm halo} = \frac{\mathcal{F}(S_n | \delta_m, S_m)}{\mathcal{F}(S_n)}.
\end{equation}

In this work, we are interested in the halo bias $b(k)$, which relates the halo overdensity to the matter overdensity,
\begin{align}
\delta^{\rm halo}(k) = b(k)\delta(k).
\end{align}
Extracting $b(k)$ from the smoothed quantities used in the excursion set formalism will require a little work, but the basic idea is that quantities smoothed on scales $R$ carry information about wavenumbers $k$ of order $R^{-1}$. So we will expand the right-hand side of Eq.\ (\ref{eqn:halooverdensity}) in $\delta_m$ and identify the coefficient of the first order term. The haloes are collapsed on scales $S_n$, but we are interested in their clustering on the much larger scales associated with $S_m$. This coefficient can then be massaged to extract $b(k)$. 

As is, the expression for the probability density, \ec{trajprobdens}, is not very useful. One of the insights of Maggiore and Riotto was to manipulate the terms so that the right side of that equation contains sums of $p$-point functions. The trick is to use the integral representation of the Dirac delta function,
\begin{align}
\delta^D(x) = \int_{-\infty}^{\infty}\frac{d\lambda}{(2\pi)}e^{-i\lambda x},
\end{align}
so that 
\begin{align}
W(\delta_0; \delta_1, ... ,\delta_n ; S_n)   =  \int_{-\infty}^{\infty}\mathcal{D}\lambda\, 
e^{i\sum_{i = 1}^{n}\lambda_{i}\delta_{i}}\langle e^{-i\sum_{i = 1}^{n}\lambda_{i}\delta(S_{i})}\rangle ,
\eql{wint}
\end{align}
where we have defined
\begin{align}
\int_{-\infty}^{\infty}\mathcal{D}\lambda \equiv & \int_{-\infty}^{\infty}\frac{d\lambda_{1}}{(2\pi)}\ldots \frac{d\lambda_n}{(2\pi)}.
\end{align}
The expectation value, $\langle e^{-i\sum_{i = 1}^{n}\lambda_{i}\delta(S_{i})}\rangle $ can be rewritten as
\begin{align} \eql{expnpoint}
\langle &e^{-i\sum_{i = 1}^{n}\lambda_{i}\delta(S_{i})}\rangle \\\nonumber
 & \quad=  \exp\left[\sum_{p = 2}^{\infty}\frac{(-i)^p}{p!}\sum^n_{j_1,...,j_p = 1}\lambda_{j_1}...\lambda_{j_p}\langle \delta(S_{j1})..\delta(S_{j_p})\rangle_c\right]
\end{align}
where $\langle \delta(S_{j1})..\delta(S_{j_p})\rangle_c$ is the connected $p$-point function. In the Gaussian case, the $p$-pt functions with $p>2$ vanish, and the probability density reduces to the limit:
\begin{align}\label{eqn:Wgm}
W^{\rm g}_{0n}   = 
 \int_{-\infty}^{\infty}\mathcal{D}\lambda\; e^{ i\sum_{i = 1}^{n}\lambda_i\delta_i - \frac{1}{2}\sum_{i,j = 1}^n \lambda_i\lambda_j \langle \delta_{i}\delta_{j} \rangle},
 \end{align}
 where the subscript denotes the initial and final times of the trajectories; the superscript identifies this as the Gaussian limit; and we suppress the dependence on the $\delta_1,\ldots,\delta_{n-1}$. We will also have occasion to invoke trajectories with starting time $S_m$ instead of $S_0$; these will be drawn from the distribution $W^{\rm g}_{mn}$ and will be integrated to obtain the probability
\begin{align}
\Pi^{\rm g}_{mn} \equiv  \Pi^{\rm g}(\delta_m, S_m; \delta_n; S_n).
\end{align}
Note that the Gaussian expression in \ec{Wgm}  satisfies the identity
\begin{align}\label{eqn:derividentity}
 \partial_{k_1}\partial_{k_2}...\partial_{k_n}W^{\rm g}_{0n}  &
 =(i)^{n}\int_{-\infty}^{\infty}\mathcal{D}\lambda\;\lambda_{k_1}\ldots \lambda_{k_n}\\\nonumber
  &\qquad\times e^{i\sum_{i = 1}^{n}\lambda_{i}\delta_{i}} e^{- \frac{1}{2}\sum_{i,j = 1}^{n} \langle \delta_{i}\delta_{j} \rangle\lambda_{i}\lambda_{j}},
\end{align}
where $\partial_i\equiv \partial/\partial\delta_i$.
Using this identity in \eec{wint}{expnpoint} leads to a general expression
for the probability density in terms of the $p$-point functions and derivatives of the Gaussian density.
\begin{align}\label{eqn:Wzng}
W_{0n}&\equiv
W(\delta_0; \delta_1, ... ,\delta_n ; S_n)  \\\nonumber
& = \exp\left[\sum_{p = 3}^{\infty}\frac{(-1)^p}{p!}\sum^n_{j_1,...,j_p = 1}\langle \delta_{j_1}..\delta_{j_p}\rangle_c\partial_{j_1}...\partial_{j_p}\right] W^{\rm g}_{0n}.
\end{align}

The effects of non-Gaussianity primarily arise from the non-zero three-point function, so we will drop all higher order $p$-point functions (and all terms quadratic and higher in the 3-point function). 
In this case, the probability density reduces to
\begin{align}\label{eqn:Wng}
W_{0n} 
 \approx  \left[1-\frac{1}{6}\sum^n_{i,j, k = 1}\langle \delta_{i}\delta_{j}\delta_{k}\rangle_c\partial_{i}\partial_{j}\partial_{k}\right] W^{\rm g}_{0n}.
\end{align}
We are never interested in the raw probability density described by \ec{Wng}. Rather, we are interested in the probability of arriving at a point $\delta_n$ at time $S_n$ or in the collapse fraction. Thus, rather than evaluate Eq.\ (\ref{eqn:Wng}) directly, we will integrate over intermediary values of $\delta_i$ to form $\Pi$ and the (observable) functions that can be constructed from it.

One well-known limit is the case when the barrier is constant (all $B$'s in \ec{probabilityofdelta} are the same) and the window function in \ec{defdel} is a top-hat in $k$-space. The top hat filter in $k$-space means that moving to larger $S$ corresponds to adding more Fourier modes, and since each Fourier mode is independent, this filter leads to a Markovian trajectory. Throughout we stick to the top-hat in $k$-space filter. In this limit of Gaussian perturbations (g) and constant barrier (cb), the probability of reaching $\delta_n$ at $S_n$ starting from $\delta=0$ at $S=0$ is
\be
\Pi^{\rm g,cb}(\delta_n,S_n) = \frac{1}{\sqrt{2\pi S_n}} \left( e^{-\delta_n^2/2S_n} - e^{-(2B-\delta_n)^2/2S_n}\right)\eql{gmcb}
.\ee
The observables then are easily computed: the collapse fraction via \ec{collapse}, 
\be
f_{\rm coll}^{\rm g,cb}(S_n) = {\rm erfc}\left( \frac{B}{\sqrt{2S_n}}\right)
\ee
and then the formation rate from \ec{form},
\be
\mathcal{F}^{\rm g,cb}(S_n) = \frac{B}{\sqrt{2\pi S_n^3}} e^{-B/\sqrt{2S_n}}
.\ee
For the remainder of this work, we will stick with the top hat in $k$-space filter, but will work to extend the collapse fraction and formation rate to allow for a moving barrier and a non-zero 3-point function. In our notation, this means we will work to generalize and drop the $^{\rm g}$ and $^{\rm cb}$ superscripts.\footnote{Note that the use of a top hat filter in $k-$space means that in the Gaussian limit the trajectories become Markovian, and thus the limit we refer to as the Gaussian limit is technically the Gaussian-Markovian limit. The effects of smoothing with a more realistic filter are relatively straight forward to include \cite{Maggiore:2009rv, Paranjape:2011ak}, see \cite{Ma:2010ep, Paranjape:2011ak, Musso:2012ch} for a study of their effects on the halo bias.}

%
\section{Unconditional crossing rate}\label{sec:uncondcross}
%
Starting from \ec{Wng}, we compute the probability of reaching $\delta_n$ at time $S_n$ allowing for any possible set of intermediate $\delta$'s.
To carry out the calculation beyond the Gaussian, constant-barrier limit of \ec{gmcb}, we first need to specify how the barrier depends on scale.
Following \cite{DeSimone:2011dn}, we can specify any barrier by its Taylor expansion about the endpoint
\begin{align}\label{eqn:barrierexpansion}
B_i = B_{n}+\sum_{p = 1}^{\infty}\frac{B^{(p)}_n}{p!}(S_i - S_n)^{p}.
\end{align}
A simple way to account for the moving barrier when computing $\Pi$ in \ec{probabilityofdelta}
is to shift integration variables
\begin{align}\label{eqn:integralrotation}
\delta_{i} \to \delta' \equiv\delta_i - \sum_{p = 1}^{\infty}\frac{B^{(p)}_n}{p!}(S_i - S_n)^{p}.
\end{align}
Then all the upper limits in the integrals leading to $\Pi$ become identical, equal to $B_n$. The integrand (\ec{Wng}) is an operator acting on $W^{g}_{0n}$, 
which now becomes: 
\bea
W^{\rm g}_{0n} &=&
\int_{-\infty}^{\infty}\mathcal{D}\lambda\; \exp\Bigg\{ i\sum_{i = 1}^{n}\lambda_i\left[\delta_i + \sum_{p = 1}^{\infty}\frac{B^{(p)}_n}{p!}
(S_i - S_n)^{p}\right] \vs
&&-\frac{1}{2}\sum_{i,j = 1}^n \lambda_i\lambda_j \langle \delta_{i}\delta_{j} \rangle\Bigg\}.
\eea
Integrating to form $\Pi$ leads to a Gaussian and non-Gaussian term:
\bea
\Pi^{\rm g}(\delta_n,S_n) &=& \Pi^{\rm g}(\delta_n,S_n) -\frac{1}{6}\sum^n_{i,j, k = 1}\langle \delta_{i}\delta_{j}\delta_{k}\rangle_c\vs
&& \times\left[ \prod_{l = 1}^{n-1}\int_{-\infty}^{B_n}d\delta_l\right] \, \partial_{i}\partial_{j}\partial_{k}W^{\rm g}_{0n}.
\eql{pim}
\eea

\subsection{Gaussian limit}

Consider the Gaussian part of this, the first term on the right in \ec{pim}.
In the limit of a constant barrier, this is straightforward to obtain and one finds the result \ec{gmcb}. However, for a moving barrier, things are complicated by the presence of the additional factor in the exponent. To evaluate the probability distribution in this case, we expand the exponential
\bea
&& \exp\Bigg\{ i\sum_{i = 1}^{n}\lambda_i \sum_{p = 1}^{\infty}\frac{B^{(p)}_n}{p!}(S_i - S_n)^{p}\Bigg\}\\ \nonumber&&
  = 1+\sum_{i = 1}^{n}\sum_{p = 1}^{\infty}\frac{B^{(p)}_n}{p!}(S_i - S_n)^{p}\partial_{i}\\\nonumber
  && +\frac{1}{2}\sum_{i ,j= 1}^{n}\sum_{p,q = 1}^{\infty}\frac{B^{(p)}_nB^{(q)}_n}{p!q!}(S_i - S_n)^{p}(S_j - S_n)^{q}\partial_{i}\partial_j+\ldots,
\eea
where the derivatives are understood to be acting on the constant barrier limit of $W$. In the Sheth-Tormen approximation \cite{Lam:2009nb, Sheth:2001dp}, one approximates $(S_n - S_i)^{p-1}\simeq S_n^{p-1}$ while simultaneously truncating the sum at $p = 5$ terms, as discussed in \cite{DeSimone:2011dn}. One can then evaluate the unconditional probability in the presence of a moving barrier in the Gaussian limit. This is given by \cite{DeSimone:2010mu,DeSimone:2011dn}
\begin{align}\nonumber\label{eqn:Piuncond}
\Pi^{\rm g}(\delta_n, S_n) \simeq &\,\Pi^{\rm g,cb}(\delta_n,S_n)\\ \nonumber
&+ 2\frac{B_n - \delta_n}{\sqrt{2\pi S_n^3}} e^{-\frac{(2B_n -\delta_n)^2}{2S_n }}\mathcal{P}_{0n} \\
& -2\frac{(B_n - \delta_n)^2}{\sqrt{2\pi S_n ^5}} e^{-\frac{(2B_n -\delta_n )^2}{2 S_n }}\mathcal{P}_{0n}^2+\ldots,
\end{align}
where
\begin{align}\label{eqn:pdef}
\mathcal{P}_{mn} \equiv\sum_{p = 1}^{5}\frac{(S_m - S_n)^p}{p!}B_n^{(p)},
\end{align}
and the $\ldots$ refer to terms higher order in an expansion in $\mathcal{P}_{0n}$, and we are assuming that the barrier is varying only slowly in time, $S_n$.
It is straightforward to integrate this over $\delta_n$ to obtain the collapse fraction,
\begin{align}\nonumber\eql{fcollnst}
f_{\rm coll}^{\rm g}(S_n) =&  {\rm erfc}\[\frac{B_{n} - \delta_m}{\sqrt{2S_n}}\] \\ \nonumber& -
\sqrt{\frac{2S_n}{\pi}} \mathcal{P}_{0n}
\left[ e^{-B_n^2/2S_n} - \sqrt{\frac{\pi B_n^2}{2S_n}}  {\rm erfc}\[\frac{B_n}{\sqrt{2S_n}}\]\right]\\\nonumber
&+\frac{\mathcal{P}_{0n}^2}{S_n^2}\left(B_n^2+S_n\right) \left(\text{erfc}\[-\frac{B_n}{\sqrt{2}
   \sqrt{S_n}}\]-2\right)\\ &+\frac{\mathcal{P}_{0n}^2}{S_n^2} \sqrt{\frac{2}{\pi }} \sqrt{S_n} B_n e^{-\frac{B_n^2}{2
   S_n}}\ldots
\end{align}
To obtain the formation rate, or the mass function, there are two ways of proceeding. One could directly differentiate the expression in \ec{fcollnst} with respect to $S_n$ to obtain the formation rate. Alternatively, since the  Gaussian-Markovian probability distribution satisfies the Fokker-Planck equation, the formation rate can be found directly by differentiating the probability distribution,
\begin{align}\eql{fpcross}
\mathcal{F}^{\rm g}(S_n) = & -\frac{1}{2}\left.\frac{\partial \Pi}{\partial \delta}\right|_{\delta = B_n}.
\end{align}
Of course, these two prescriptions should lead to the same result. However, due to the rather crude approximation that leads to Eq.\ (\ref{eqn:Piuncond}), this expression no longer satisfies the Fokker-Planck equation. This is an obvious drawback of the crude approximation we used in order to evaluate the probability distribution. In Appendix \ref{app:linearbarrier}, we demonstrate that enforcing the Fokker-Planck equation when calculating the formation rates (i.e.\ using \ec{fpcross}), rather than directly differentiating the collapse fraction, leads to results that are consistent with those obtained by an exact treatment of the probability distribution in the case of a linear barrier where no such approximation needs to be made. Thus, for the rest of this work, where possible we apply the Fokker-Planck equation.

Using \ec{fpcross} and Eq.\ (\ref{eqn:Piuncond}) we obtain the unconditional formation rate for a general barrier
\begin{align}\label{eqn:uncondcrossrateg}
\mathcal{F}^{\rm g}(S_n) = & \frac{B_n+\mathcal{P}_{0n}}{\sqrt{2\pi S_n^3}}e^{-\frac{B_n^2}{2S_n}}.
\end{align}

Eq.~\Ec{uncondcrossrateg} is a compact expression that agrees with that obtained in previous work, but as described above, it hides a subtlety in the Sheth-Tormen approximation. Had we simply differentiated \ec{fcollnst}, we would have arrived at a considerably more complicated looking result, however, numerically the difference is small.

\subsection{Non-Gaussian contribution}

To compute the non-Gaussian piece in \ec{pim}, we need to carry out the sums (integrals in the limit that $\epsilon\rightarrow0$). Maggiore and Riotto showed that the simple approximation of evaluating the 3-point function at the end point $i=j=k=n$ leads to the first term in an expansion in the parameter $S_n/B_n^2$. A useful way of evaluating the remaining integrals is to use the identity
\cite{Maggiore:2009rx, DeSimone:2010mu}
\begin{align}\label{eqn:identity}
\frac{\partial^3}{\partial B_n^3}f_{\rm coll}^{\rm g}(S_n) = &-\!\!\!\sum_{j_1,j_2,j_3 = 1}^{n}\int_{-\infty}^{B_n}d\delta_{1}\ldots\!\int_{-\infty}^{B_n}d\delta_{n}\prod_{i = 1}^{3}\partial_{j_i} W^{\rm g}_{0n}.
\end{align}
This is almost identical to the sum and integrals in the last term in \ec{pim}, with the exception that the right hand side in \ec{identity} also includes an integral over $\delta_n$. To use the identity then, we can integrate \ec{pim} over $\delta_n$ (which we need to do anyway in order to compute the collapse fraction). This leaves
\be
f_{\rm coll}^{\rm NG} = \frac{1}{6} \langle \delta_n^3 \rangle  \, \frac{\partial^3}{\partial B_n^3} f_{\rm coll}^{\rm g}(S_n)
.\ee
To obtain the crossing rate, we differentiate this with respect to $S_n$. The derivative acting on the three point function is straightforward. To evaluate the derivative acting on $f_{\rm coll}^{\rm g}$, we pull the derivative all the way through: $\int d\delta \partial\Pi(\delta;S_n)/\partial S_n$; use the Fokker-Planck equation and then integrate by parts so that the non-Gaussian part of the collapse fraction is 
\begin{align}\nonumber
\mathcal{F}^{\rm NG}(S_n) = & -\left.\frac{1}{2}\left(1-\frac{1}{6}S_n^2\mathcal{S}_{3}\frac{\partial^3}{\partial B_n^3}\right)\frac{\partial \Pi^{\rm g}}{\partial \delta}\right|_{\delta = B_n}\\
&  - \frac{1}{6}(2 S_n \mathcal{S}_{3} +S_n^{2}\mathcal{S}_{3} ')\frac{\partial^3}{\partial B_n^3}f_{\rm coll}^{\rm g}(S_n).
\end{align}
We have defined
\begin{align}
\mathcal{S}_{3} \equiv \frac{\langle \delta_n^3 \rangle}{S_n^2}
\end{align}
and denoted derivatives with respect to $S_n$ by primes, $'$. Using \ec{fcollnst} and \ec{Piuncond} we find that the full unconditional crossing rate to leading order in the three point function is given by
\begin{align}\label{eqn:uncondcrossrate}
\mathcal{F}(S_n) = &  \frac{B_n+\mathcal{P}_{0n}}{\sqrt{2\pi S_n^3}}e^{-\frac{B_n^2}{2S_n}} \\ \nonumber
& +\frac{\mathcal{S}_{3} e^{-\frac{B_n^2}{2S_n}}}{6\sqrt{2\pi S_n^5}}\bigg[\mathcal{P}_{0n}B_n \left(B_n^2+S_n\right)-10 B_n^2 S_n\\ \nonumber
& +B_n^4-8 \mathcal{P}_{0n}^2 S_n+7 S_n^2\bigg]\\\nonumber
&+\frac{S_n^2\mathcal{S}'_{3}e^{-\frac{B_n^2}{2S_n}} }{3\sqrt{2\pi S_n^5}}\[B_n^2 - B_n\mathcal{P}_{0n}-S_n+2\mathcal{P}_{0n}^2\].
\end{align}

%
\section{Conditional crossing rate}\label{sec:condcross}
%

In order to evaluate the halo overdensity, Eq.\ (\ref{eqn:halooverdensity}), we also need the conditional crossing rate. The major difference between evaluating the probability distribution in this case (Eq.\ (\ref{eqn:probabilityofdeltacond})) and the previous case (Eq. (\ref{eqn:probabilityofdelta})) is that the intermediate point on the trajectory denoted by $\delta_m$ is not integrated over. Using the relation between $W$ and $W^{\rm g}$ in \ec{Wng}, the starting point for the conditional crossing rate calculation is then
\bea
&&\Pi(\delta_n,S_n | \delta_m, S_m)\simeq \Bigg[  
\int_{-\infty}^{B_1}\!\!d\delta_{1}\ldots \widehat{d\delta}_{m}\ldots\int_{-\infty}^{B_{n-1}}\!\!d\delta_{n - 1} 
\vs
&&\qquad\times  \left( 1 -\frac{1}{6}\sum^n_{i,j, k = 1}\langle \delta_{i}\delta_{j}\delta_{k}\rangle_c\partial_{i}\partial_{j}\partial_{k}
 \right) W_{0n}^{\rm g}  \Bigg]
\vs
&&\qquad\qquad\times\Bigg[\int_{-\infty}^{B_1}\!\!d\delta_{1}\ldots\int_{-\infty}^{B_{m-1}}\!\!d\delta_{m - 1} \vs
&&\qquad\quad\quad\times
\left( 1 -\frac{1}{6}\sum^m_{i,j, k = 1}\langle \delta_{i}\delta_{j}\delta_{k}\rangle_c\partial_{i}\partial_{j}\partial_{k}
 \right) W_{0m}^{\rm g} 
 \Bigg]^{-1}.\eql{constart}
\eea

\Sfig{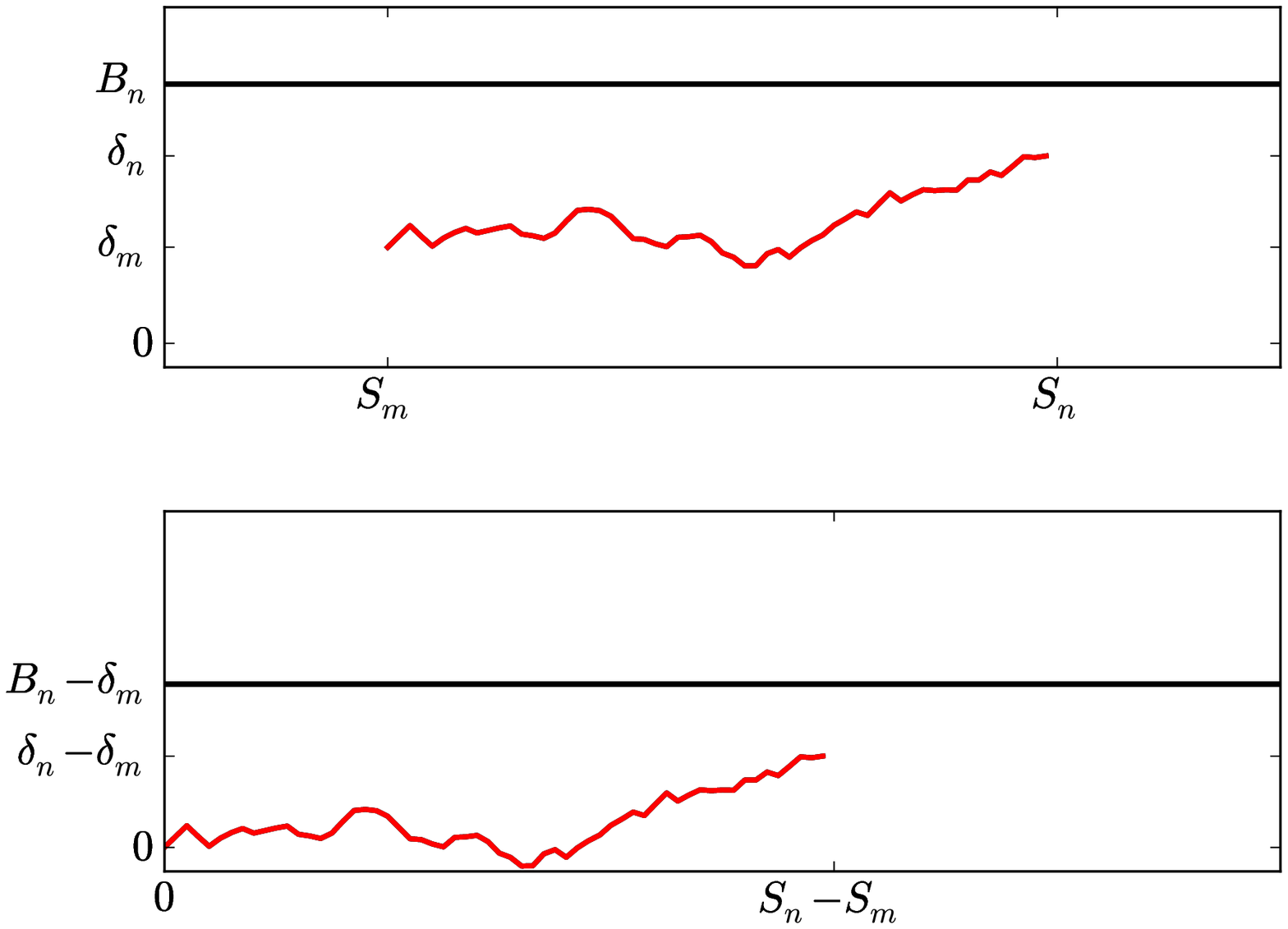}{Top panel shows a trajectory from large scale $S_m$ to small scale $S_n$. In the Gaussian, Markovian limit, the probability for this trajectory does not depend on prior steps, so is equivalent to the trajectory depicted in the bottom panel with the shifts enumerated in \ec{shift}.}

\subsection{Gaussian contribution}\label{sec:cong}

We first consider the Gaussian piece of \ec{constart} by dropping all terms with 3-point functions in them. In this limit, there is a simplification due to the Markovian nature of the trajectories. Namely, the probabilities do not depend on the paths chosen but simply on the end points. So $W_{0n}^{\rm g}$
in the numerator can be replaced with $W_{0m}^{\rm g} W_{mn}^{\rm g}$. Then all the integrals over $\delta_1,\delta_2,\ldots,\delta_{m-1}$ are identical in the numerator and denominator, leading to
\be
\Pi^{\rm g}(\delta_n,S_n | \delta_m, S_m) =
\int_{-\infty}^{B_{m+1}}\!\!d\delta_{m+1}\ldots \int_{-\infty}^{B_{n-1}}\!\!d\delta_{n - 1} W_{mn}^{\rm g}.\ee
As depicted in Fig.~\rf{traj}, this expression is identical to the unconditional Gaussian Markovian probability (which we evaluated in \ec{Piuncond}) as long as we make the substitutions
\bea
\delta_n &\rightarrow &\delta_n-\delta_m\vs
B_n &\rightarrow & B_n - \delta_m \vs
S_n &\rightarrow& S_n-S_m\eql{shift}
.\eea

For example, the first term in \ec{Piuncond} -- the constant barrier piece -- becomes
\bea
\Pi^{\rm g,cb}(\delta_n,S_n\vert \delta_m, S_m) &=& \frac{1}{\sqrt{2\pi [S_n-S_m]}} \vs
\times \big( e^{-[\delta_n-\delta_m]^2/2[S_n-S_m]} &-& e^{-(2B-\delta_n-\delta_m)^2/2[S_n-S_m]}\big)
.\vs\eea

\subsection{Non-Gaussian contribution}

To evaluate the non-Gaussian corrections to the conditional collapse rate, we again rely on the approximation of considering the three-point functions only at their end points. In the case of unconditional collapse, this translated to setting $\langle \delta_i\delta_j\delta_k\rangle \rightarrow \langle \delta_n\delta_n\delta_n\rangle$, thereby simplifying the sum over $i,j,k$. Here, too, we set all three-point functions to their values at the end point but when $i<m$, the end point is time $S_m$, not $S_n$.
For example, 
\begin{align}\label{eqn:correlatorapprx}
\sum_{i,j = m+1}^{n}\sum_{k = 1}^{m -1}\langle \hdelta_i\hdelta_j\hdelta_k\rangle_c \approx \langle \hdelta_m\hdelta_n^2\rangle_c\sum_{i,k = m+1}^{n}\sum_{k = 1}^{m -1}.
\end{align}
and working to linear order in the long wavelength mode, $\delta_m$, we can write (see also \cite{D'Aloisio:2011mn})
\begin{align}\nonumber\label{eqn:lininm}
\sum_{i,j,k = 1}^{n}&\langle \hdelta_i\hdelta_j\hdelta_k\rangle_c\partial_{i}\partial_{j}\partial_{k}W^{\rm g}_{0, n}  \\\nonumber
& =3\langle \hdelta_m\hdelta_n^2\rangle_c\sum_{k = 1}^{m -1}\sum_{i,j = m+1}^{n}\partial_{k} W^{\rm g}_{0, m}\partial_{i}\partial_{j}W^{\rm g}_{m, n}\\ \nonumber
& +
3\langle \hdelta_n^2\hdelta_m\rangle_c\sum_{i,j = m+1}^{n} \partial_{m} (W^{\rm g}_{0, m}\partial_{i}\partial_{j}W^{\rm g}_{m, n})\\ & 
+\langle \hdelta_n^3\rangle_cW^{\rm g}_{0, m}\sum_{i,j,k = m+1}^{n}\partial_{i}\partial_{j}\partial_{k}W^{\rm g}_{m, n}.
\end{align}
Rotating the integrals in the analogous manner as Eqs.\ (\ref{eqn:barrierexpansion})-(\ref{eqn:integralrotation}) and making use of identities such as Eq.\ (\ref{eqn:identity}) as well as
\begin{align}
\frac{\partial^N}{\partial B_n^N}\Pi_{0m}^{\rm g} = & \sum_{j_i,..,j_N = 1}^{m-1}\int_{-\infty}^{B_n}d\delta_{1}\ldots \int_{-\infty}^{B_n}d\delta_{m-1}\prod_{i = 1}^{N}\partial_{j_i}W^{\rm g}_{0, m},
\end{align}
we find that the non-Gaussian part of the conditional collapse fraction can be written
\begin{widetext}
\begin{align}
f_{\rm coll}^{\rm NG}(S_n | \delta_m, S_m) = & 
-\frac{1}{6}\Bigg[\(\langle \hdelta_n^3\rangle_c \frac{\partial^3}{\partial B_n^3}+
3\langle \hdelta_n^2\hdelta_m\rangle_c \frac{\partial^2}{\partial B_n^2}\frac{\partial}{\partial \delta_m}\)
+ 3\langle \hdelta_m\hdelta_n^2\rangle_c\left(\frac{\partial}{\partial B_n}+\frac{\partial}{\partial \delta_{m}}\right)\ln\Pi_{0,m}^{\rm g}\frac{\partial^2}{\partial B_n^2}
\Bigg] f^{\rm g}_{\rm coll}(S_n|\delta_m, S_m),
\end{align}
where $f^{\rm g}_{\rm coll}(S_n|\delta_m, S_m)$ is the conditional collapse fraction in the Gaussian limit, which follows from the results of \S\ref{sec:cong}.

Taking the derivatives and working in the limit $S_m\to 0$ with $B_n \gg \delta_m$, we find that the non-Gaussian contribution to the conditional collapse fraction is
\begin{align}\label{eqn:concoll}\nonumber
& f_{\rm coll}^{\rm NG}(S_n| \delta_m, S_m) =  
-\frac{1}{6}\Bigg[2(\langle \hdelta_n^3\rangle_c -3\langle \hdelta_n^2\delta_m \rangle_c)\frac{e^{-\frac{(B_{n} - \delta_m)^2}{2(S_n - S_m)}}}{\sqrt{2\pi (S_n - S_m)^3}}\Bigg(1-\frac{(B_{n}-\delta_m)^2}{(S_n - S_m)}+\frac{(B_{n}-\delta_m)}{(S_n-S_m)}\mathcal{P}_{nm}-2\frac{\mathcal{P}_{nm}^2}{(S_n-S_m)}\Bigg)\\ 
& +3\langle \hdelta_m\hdelta_n^2\rangle_c \frac{\delta_m}{S_m}\left(\frac{2\exp\[-\frac{(B_{n} - \delta_m)^2}{2(S_n - S_m)}\]}{\sqrt{2\pi (S_n - S_m)^3}}\( \mathcal{P}_{mn} -(B_n - \delta_m)\)-\frac{2}{(S_n - S_m)^2}{\rm erfc}\[\frac{B_{n} - \delta_m}{\sqrt{2(S_n - S_m)}}\] \mathcal{P}_{mn}^2\right)\Bigg].
\end{align}
The key result here is the appearance of the combination
\bea\label{eqn:finiteinmtozero}
\langle \hdelta_m\hdelta_n^2\rangle_c\left(\frac{\partial}{\partial B_n}+\frac{\partial}{\partial \delta_{m}}\right)\ln\Pi_{0,m}^{\rm g} \approx \langle \hdelta_m\hdelta_n^2\rangle_c\frac{\delta_m}{S_m},
\eea
which multiplies the last line in \ec{concoll}. The appearance of the terms proportional to $\langle \hdelta_m\hdelta_n^2\rangle_c$ in  Eq.\ (\ref{eqn:concoll})  encode the fact that the conditional probability of collapse now depends on the details of the correlation between long and short wavelength fluctuations. We will see that these correlations give rise to a scale dependent bias. In order to extract the bias we will eventually correlate the expression in Eq.\ (\ref{eqn:concoll}) (to linear order in $\delta_m$) with the underlying density fluctuations smoothed on the scale $R_m$, before taking the limit that $S_m \to 0$. While the quantity in Eq. (\ref{eqn:finiteinmtozero}) remains finite after this procedure, the terms in the first line of Eq.\ (\ref{eqn:concoll}) proportional to $\langle \hdelta_m\hdelta_n^2\rangle_c$ vanish, and we thus drop them in what follows.

We can then evaluate the conditional crossing rate, which is the conditional rate at which objects form at time $S_n$, given that they had an overdensity $\delta_m$ at time $S_m$. As we have stressed above, we do not proceed by simply differentiating Eq.\ (\ref{eqn:concoll}) with respect to $S_n$. Rather, we write the non-Gaussian contribution to the conditional crossing rate as
\begin{align}\nonumber
 \mathcal{F}^{\rm mb, NG}&(S_n|\delta_m, S_m)  =  \frac{\partial f_{\rm coll}^{\rm NG}(S_n| \delta_m, S_m)}{\partial S_n}\\\nonumber
 = & \frac{1}{6}\Bigg[S_n^2\mathcal{S}_3 \frac{\partial^3}{\partial B_n^3}
+ 3\sqrt{S_m}S_n\mathcal{S}_{n^2 m}\left(\frac{\partial}{\partial B_n}+\frac{\partial}{\partial \delta_{m}}\right)\ln\Pi_{0,m}^{\rm g}\frac{\partial^2}{\partial B_n^2}
\Bigg]\frac{1}{2}\left.\frac{\partial \Pi^{\rm g}_{m,n}}{\partial \delta_n}\right|_{\delta_n = B_n}\\ 
& - \frac{1}{6}\Bigg[(2 S_n \mathcal{S}_3+S_n^2 S'_{3}) \frac{\partial^3}{\partial B_n^3}+
 + 3\sqrt{S_m}(\mathcal{S}_{n^2 m}+S_n \mathcal{S}'_{n^2 m})\left(\frac{\partial}{\partial B_n}+\frac{\partial}{\partial \delta_{m}}\right)\ln\Pi_{0,m}^{\rm g}\frac{\partial^2}{\partial B_n^2}
\Bigg] f^{\rm g}_{\rm coll}(S_n|\delta_m, S_m),
\end{align}
where we have made use of the Fokker-Planck equation in the second line and primes denote derivatives with respect to $S_n$. We have dropped the terms which vanish in the $S_m\to0$ limit as described above and defined 
\begin{align}
\mathcal{S}_{n^2 m} = \frac{\langle \delta^{2}_n \delta_m\rangle_c}{S_n \sqrt{S_m}}.
\end{align}
We then evaluate to find the conditional crossing rate
\begin{align}\label{eqn:condcrossrate}\nonumber
 \mathcal{F}^{\rm mb, NG}(S_n|\delta_m, S_m) = &    -\frac{1}{3} \mathcal{S}'_{n^3}\frac{e^{-\frac{(B_{n} - \delta_m)^2}{2(S_n - S_m)}}}{\sqrt{2\pi (S_n - S_m)}}\[(S_n - S_m)-(B_{n}-\delta_m)^2+(B_{n}-\delta_m)\mathcal{P}_{nm}-2\mathcal{P}_{nm}^2\]\\\nonumber
& +\frac{\mathcal{S}_{n^3}}{6}\frac{e^{-\frac{(B_n - \delta_m)^2}{2 (S_n - S_m)}}}{\sqrt{2\pi (S_n - S_m)^5}} \bigg[\mathcal{P}_{mn}(B_n - \delta_m) \left((B_n-\delta_m)^2+(S_n - S_m)\right)-10 (B_n - \delta_n)^2 (S_n - S_m)
\\ \nonumber
& +(B_n-\delta_m)^4 -8 \mathcal{P}_{mn}^2 (S_n - S_m)+7 (S_n - S_m)^2\bigg]\\\nonumber
  &+  \sqrt{S_m}\mathcal{S}'_{n^2 m}\frac{\delta_m}{S_m}\Bigg[\frac{e^{-\frac{(B_{n} - \delta_m)^2}{2(S_n - S_m)}}}{\sqrt{2\pi (S_n - S_m)}}\( \mathcal{P}_{mn} -(B_n - \delta_m)\)-\frac{{\rm erfc}\[\frac{B_{n} - \delta_m}{\sqrt{2(S_n - S_m)}}\] }{(S_n - S_m)} \mathcal{P}_{mn}^2\Bigg]\\\nonumber
&-\sqrt{S_m}\mathcal{S}_{n^2 m}\frac{\delta_m}{S_m}\frac{e^{-\frac{(B_{n} - \delta_m)^2}{2(S_n - S_m)}}}{\sqrt{2\pi (S_n - S_m)^3}}\Bigg[\frac{(B_n - \delta_m)^2}{S_n - S_m}(B_n - \delta_m+\mathcal{P}_{mn}) - 3\mathcal{P}_{mn} - (B_n - \delta_m)\\ & +2\sqrt{\frac{2\pi}{S_n-S_m}} e^{\frac{(B_{n} - \delta_m)^2}{2(S_n - S_m)}}\mathcal{P}_{mn}^2{\rm erfc}\[\frac{B_n-\delta_m}{\sqrt{2(S_n - S_m)}}\]\Bigg].
\end{align}

%
\section{Halo bias}\label{sec:halobias}
%

We now have  all the ingredients with which to extract the halo bias. Taking the ratio of Eq.\ (\ref{eqn:condcrossrate}) and Eq.\ (\ref{eqn:uncondcrossrate}), and expanding to linear order in $\delta_m$, we find that the relationship between the halo overdensity and the matter overdensity is
\begin{align}\nonumber\label{eqn:diasrealspace}
\delta^{\rm halo}_{m} = & \(\frac{B_n}{S_n} - \frac{1}{B_n + \mathcal{P}_{0n}}\)\delta_m \\\nonumber
& - \frac{1}{6}\frac{\mathcal{S}_{n^3}}{(B_n + \mathcal{P}_{0n})^2}\Bigg(20 \mathcal{P}_{0n}B_n+10 B_n^2-9 \mathcal{P}_{0n}^2-3 \frac{B_n^2}{S_n} \left(B_n+\mathcal{P}_{0n}\right){}^2+7 S_n
   \Bigg)\delta_m\\\nonumber
  &  -\frac{S_n\mathcal{S}'_{n^3}}{3(B_n+\mathcal{P}_{0n})^2}\Bigg( (B_n -\mathcal{P}_{0n})(B_n+3\mathcal{P}_{0n})+S_n\Bigg)\delta_m\\
\nonumber
&+\sqrt{S_m}\mathcal{S}_{n^2 m}\Bigg(B_n\(\frac{B_n}{S_n}- \frac{1}{B_n+\mathcal{P}_{0n}}\)- \frac{3\mathcal{P}_{0n} }{B_n+\mathcal{P}_{0n}}+2\sqrt{\frac{2\pi}{S_n}} \frac{e^{\frac{B_{n}^2}{2S_n }}\mathcal{P}_{0n}^2}{B_n+\mathcal{P}_{0n}}{\rm erfc}\[\frac{B_n}{\sqrt{2S_n}}\]
     \Bigg)\frac{\delta_m}{2S_m}\\ 
 & +   S_n\sqrt{S_m}\mathcal{S}'_{n^2 m}\Bigg( 1-\frac{2\mathcal{P}_{0n}}{B_n+\mathcal{P}_{0n}}+\sqrt{\frac{2\pi}{ S_n}}\frac{ \mathcal{P}_{0n}^2}{B_n+\mathcal{P}_{0n}}{\rm erfc}\[\frac{B_{n}}{\sqrt{2S_n}}\]\Bigg)\frac{\delta_m}{S_m}
\end{align}
\end{widetext}
We can immediately read off from the first line here the well-known~\cite{Sheth:2001dp} Gaussian bias in initial Lagrangian space:
\be
b^{\rm g} = \frac{B_n}{S_n} - \frac{1}{B_n + \mathcal{P}_{0n}  }.
\ee
The coefficients of $\delta_m$ in the second and third lines do not depend on the large scale $m$ and so represent the scale independent contribution to the bias due to non-Gaussian correlations between the initial fluctuations. The result presented here is slightly different from that reported by \cite{DeSimone:2011dn} who do not employ the Fokker-Planck equation. 
The last two lines contain terms wherein the coefficients of $\delta_m$ depend on the scale $m$, so the bias from these terms will be scale-dependent. Note that they are proportional to the three-point function so vanish in the Gaussian limit. Also, notice that the effects of the moving barrier are encoded in the $\mathcal{P}_{0n}$'s here. These terms then give the full contribution to the scale-dependent bias from non-Gaussianity accounting for ellipsoidal collapse.
Care is required to extract the exact form of the scale dependent bias,  because the bias is typically defined in Fourier space, while we have worked with real space fluctuations smoothed over various scales.

%
\section{The scale dependent bias from excursion sets}\label{sec:scaledepbias}
%

We can now extract the scale dependent bias from our excursion set result. For simplicity, we begin with the case of spherical collapse where the barrier is constant, $B(S_n) = \delta_c$. Focusing on the scale dependent parts of Eq. (\ref{eqn:diasrealspace}), and assuming a spherical barrier so that $\mathcal{P}_{0n}=0$, we have
\begin{align}\label{eqn:biasrealsphcol}
\delta^{\rm halo}_{m} = \sqrt{S_m}S_n\mathcal{S}'_{n^2 m}\frac{\delta_m}{S_m}+\frac{\sqrt{S_m}\mathcal{S}_{n^2 m}}{2}\Bigg(\frac{\delta_c^2}{S_n} -1 \Bigg)\frac{\delta_m}{S_m}.
\end{align} 
The problem is to transform this relation, between the real space halo overdensity smoothed on a large scale $S_m$ with the real space matter overdensity smoothed on the same scale, into a Fourier space relationship:
\begin{align}\label{eqn:biasdef}
\delta^{\rm halo}(\vec k) = b(k)\delta(\vec k).
\end{align}

In terms of the Fourier space fields, the relationship between the halo field and the matter field, as derived in the excursion set, Eq.\ (\ref{eqn:biasrealsphcol}) becomes
\begin{align}\label{eqn:scaledepbiaseqn}
 \int \frac{d^{3}k}{(2\pi)^3}\delta(\vec k) b^{\rm SD,sph}(k) W(\vec k,R_m) = & \Bigg(
 \frac{\sqrt{S_m}\mathcal{S}_{n^2m}}{2} \(\frac{\delta_c^2}{S_n}-1\)
 \vs
 +\sqrt{S_m}S_n\mathcal{S}'_{n^2m}\Bigg)  \int \frac{d^{3}k'}{(2\pi)^3} &
 \frac{\delta(\vec k') W(\vec k,R_m)}{S_m}
\end{align}
 where the $^{\rm SD,sph}$ superscript reminds us that we are dealing with the scale-dependent part of the bias only here and, for now, 
 focusing on the spherical collapse limit.
Now correlate both sides of \ec{scaledepbiaseqn} with $\delta_m$. The left hand side becomes
\begin{align}\label{eqn:oneside1}
\int_{0}^{k_{m}} \frac{d^3k}{(2\pi)^3}\, P_\delta(k) b^{\rm SD,sph}(k)  =  \int_{0}^{k_{m}}  d\ln k \;b^{\rm SD,sph}(k) \Delta_{\delta}^2(k),
\end{align}
where $\Delta^2(k)\equiv k^3P(k)/2\pi^2$, and we have defined the power spectrum in the usual way,
\begin{align}
 \langle \delta(\vec  k)\delta(\vec k')\rangle =  &  P_{\delta}(k)(2\pi)^3\delta^3(\vec k -\vec k').
\end{align}
The right-hand side becomes the set of coefficients in parenthesis in \ec{scaledepbiaseqn} multiplying
\be
\frac{1}{S_m} \int_{0}^{k_{m}} \frac{d^3k}{(2\pi)^3}\, P_\delta(k)  =  1
\ee
as the $k$ integral here is equal to $S_m$ for our chosen filter.
So the remaining equality is
\bea
\int_{0}^{k_{m}}  d\ln k \;b^{\rm SD,sph}(k)\Delta_{\delta}^2(k) &=&
\Bigg(
 \frac{\sqrt{S_m}\mathcal{S}_{n^2m}}{2} \(\frac{\delta_c^2}{S_n}-1\)\vs
&& +\sqrt{S_m}S_n\mathcal{S}'_{n^2m}\Bigg) \eql{scaledepbiaseqn2}
 .\eea

To proceed we need information about the shape of the three-point function that enters into \ec{scaledepbiaseqn2}. The primordial 3-point functions are most easily written in terms of the gravitational potential, so first recall that 
\be
\delta({\bf k}, z) =  \mathcal{M}(k,z)\Phi({\bf k})
\ee
where
\begin{align}\label{eqn:matternewt}
\mathcal{M}(k, z) = \frac{2}{3}\frac{k^{2}T(k)g(z)}{\Omega_m H_0^2(1+z)}
\end{align}
and here $T(k)$ is the matter transfer function normalized at unity as $k\to 0$, $g(z)$ is the linear growth of the gravitational potential during the matter dominated epoch, and $\Phi({\bf k})$ denotes the primordial value of the gravitational potential (hence no $z$ dependence).
A generic 3-pt function of $\Phi$ is characterized by the bispectrum,
\begin{align}
\langle \Phi({\bf k}_{1})\Phi({\bf k}_{2})\Phi({\bf k}_{3})\rangle = B_{\Phi}(k_{1}, k_{2}, k_{3})(2\pi)^{3}\delta^{3}({\bf k}_{1}+{\bf k}_{2}+{\bf k}_{3}).
\end{align}
We can then evaluate the 3-pt function of density fluctuations,
\begin{align}\label{eqn:3ptdensgen}
\langle\delta_m\delta_n^2\rangle = &\int_0^{k_m} \frac{dk_1}{k_1} \int_0^{k_n} \frac{dk_2}{k_2} \int_{-1}^{1} d\cos\theta\\ \nonumber&\times \frac{k_{1}^3}{2\pi^2} \frac{k_{2}^3}{4\pi^2}\mathcal{M}(k_1)\mathcal{M}(k_2)\mathcal{M}(k_{3}) B_{\Phi}(k_{1}, k_{2}, k_{3}),
\end{align}
where $\theta$ is the angle between $\vec{k}_{1}$ and $\vec{k}_2$ and we have assumed that $|\vec{k}_3| = |\vec k_{1}+\vec k_{2}| \leq k_n$. Following the notation of \cite{Matarrese:2008nc, Desjacques:2011mq}, we introduce the function (not to be confused with the crossing rate!)
\begin{align}\nonumber
\mathcal{F}^{3}_{n}(k) = \frac{1}{S_n\Delta^{2}_{\Phi}(k)}\int^{k_n}& \frac{k_2^2 dk_2 }{2\pi^2} \mathcal{M}(k_2)\mathcal{M}(\vec{k}_{2} - \vec{k}|)\\ & \times B_{\Phi}(k_{1}, k_{2}, |\vec{k}_{2} - \vec{k}|),
\end{align}
where $\vec k = k \hat{z}$, and $\hat z$ is an arbitrary unit vector.
In the limit $k_m \ll k_n$, Eq.\ (\ref{eqn:scaledepbiaseqn2}) then reduces to
\begin{align}\label{eqn:scaledepbiaseqn3}\nonumber
\,\int_{0}^{k_{m}}\!\!\! d\ln k\Delta_{\delta}^2(k) b^{\rm SD,sph}(k) =&  
\!\!\!\int^{k_m}\!\!\!\! d \ln k \frac{\Delta^2_{\delta}(k)}{\mathcal{M}(k)}\\
\times & \left[
\frac{1}{2} \(\frac{\delta_c^2}{S_n}-1\)\mathcal{F}^{3}_{n}(k) 
+ 
\frac{d \mathcal{F}^{3}_{n}(k)}{d\ln S_n}
\right].
\end{align}
At this stage, we have an equality between two sums of things. To see that we can equate each term in the sum, imagine that we measure these sums for two slightly different scales, $k_m$ and $k_{m}+\delta k$, then we can extract information about the contribution to the sum from scale $k_m$ by differencing these quantities. This then implies, as well as being an equality between sums, equality between each term in the sum. Then the scale dependent bias can be read off:
\begin{align}
b^{\rm SD,sph}(k) = \mathcal{M}^{-1}(k)\left[ 
\frac{1}{2} \(\frac{\delta_c^2}{S_n}-1\) \mathcal{F}^{3}_{n}(k) +\frac{d \mathcal{F}^{3}_{n}(k)}{d\ln S_n}\right],
\eql{biassph}
\end{align}
which agrees with the result derived by \cite{Desjacques:2011mq}. Note that we naturally obtain the important second term, which can be thought of arising due to the fact that a scale dependent rescaling of the variance also changes the significance interval that corresponds to a fixed mass. As we will see, this extra term vanishes for the local ansatz, but is non-zero and generally important for other bispectrum shapes \cite{Desjacques:2011mq, Desjacques:2011jb} and in particular largely ameliorates the discrepancies noted by \cite{Shandera:2010ei}. 

For the local ansatz,
\begin{align}
B_{\Phi}(k_{1}, k_{2}, k_{3}) = 2f_{\rm NL}(P(k_{1})P(k_{2})+{\rm perm.})
\end{align}
so that
\begin{align}
\mathcal{F}^{3}_{n} \approx 4f_{\rm NL}
\end{align}
a constant. Therefore only the first term in \ec{biassph} contributes leaving
\begin{align}\eql{biasspherical}
b^{\rm SD,sph,local}(k) = \frac{2f_{\rm NL}}{\mathcal{M}(k)} \(\frac{\delta_c^2}{S_n}-1\),
\end{align}
for the scale dependent bias in the spherical collapse case, 
which agrees exactly with the result of \cite{Dalal:2007cu}.

Using the same techniques, we can extract the scale dependent bias from \ec{diasrealspace} for the full ellipsoidal collapse problem. It too has pieces proportional to both $\mathcal{F}^{3}_{n}$ and $d\mathcal{F}^{3}_{n}/dS_n$. We separate these out as:
\be\eql{scaledepbiasgen}
b^{\rm SD}(k) = \frac{\mathcal{F}^{3}_{n}(k)}{2\mathcal{M}(k)} c_n  + \frac{1}{ \mathcal{M}(k)}\frac{d \mathcal{F}^{3}_{n}(k)}{d\ln S_n} d_n
\ee
where the coefficients are:
\begin{widetext}
\be
c_n \equiv 
B_{n} b^{\rm g} -\frac{3\mathcal{P}_{0n}}{B_n+\mathcal{P}_{0n} }+2 \sqrt{\frac{2\pi}{S_n}}\frac{\mathcal{P}_{0n}^2 }{(B_n+\mathcal{P}_{0n})}e^{\frac{B_n^2}{2 S_n}} \text{erfc}\left[\frac{B_n}{\sqrt{2S_n}}\right] \eql{coeff}
\ee
and
\begin{align}
d_n \equiv 1-\frac{2\mathcal{P}_{0n}}{(B_n+\mathcal{P}_{0n})}+\sqrt{\frac{2\pi}{S_n}}\frac{\mathcal{P}_{0n}^2 }{(B_n+\mathcal{P}_{0n})} e^{\frac{B_n^2}{2 S_n}} \text{erfc}\left[\frac{B_n}{\sqrt{2S_n}}\right].
\end{align}
\end{widetext}
These equations are the main results of the paper. 

\Sfig{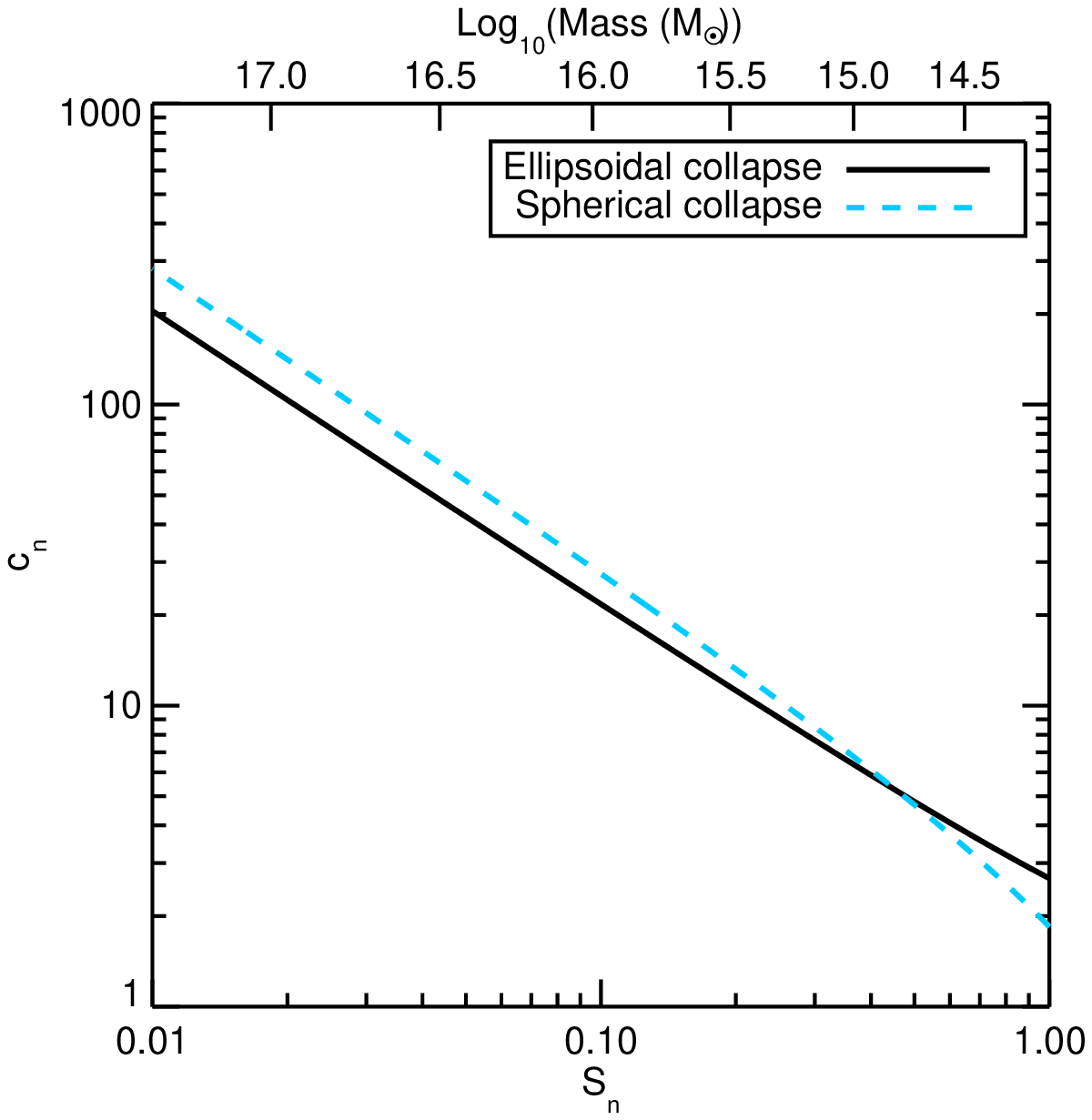}{The coefficient (as defined in \ec{scaledepbiasgen}) of the scale
  dependent bias, $c_n$, as a function of the smoothing scale
  $S_n$.  Solid (black) curve and dashed (blue) curve show the result
  in Eq. \ref{eqn:scaledepbiasgen} for the ellipsoidal
  collapse model of \cite{ShethMoTormen:2001}
  (i.e. Eq. \ref{eqn:ellcollapsebarrier}) and the spherical collapse
  model (i.e. when $B_n=\delta_c$ and $\mathcal{P}_{0n}=0$) respectively.  The upper
  x-axis shows the mass corresponding to a particular $S_n$ assuming a
  flat, $\Lambda$CDM cosmology with $\Omega_b h^2 = 0.02$, $\Omega_c
  h^2 = 0.11$, $h=0.7$ and $\sigma_{8} = 0.87$.}

\Sfig{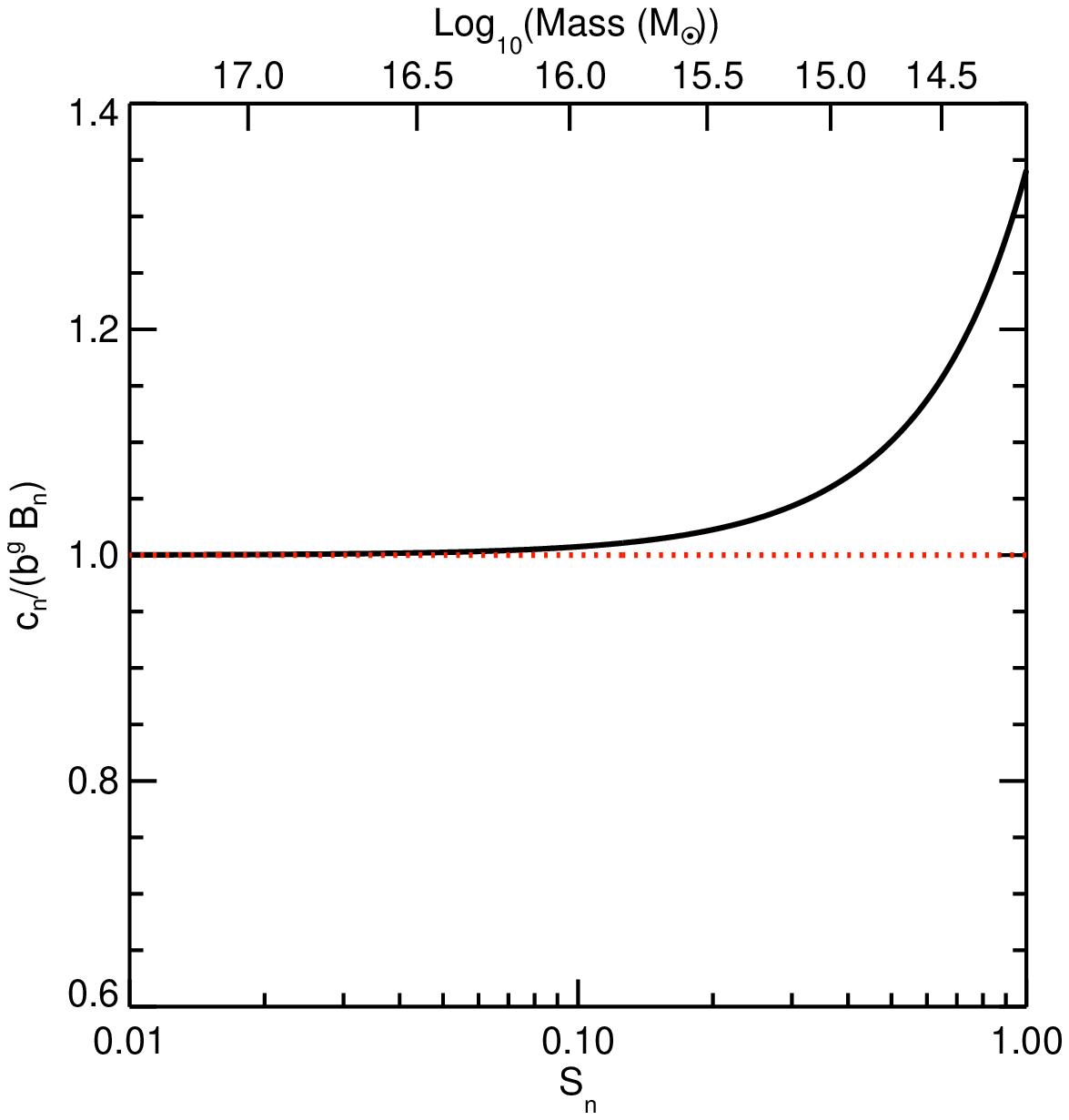}{The coefficient of the scale dependent bias,
  $c_n$, from \ec{scaledepbiasgen} for the ellipsoidal
  collapse model of \cite{ShethMoTormen:2001}, normalized to the naive
  prediction $c_n=b^{\rm g} B_n$.  The upper
  x-axis shows the mass corresponding to a particular $S_n$ assuming
  the same cosmology as in Fig. \ref{fig:bias_ellipsoidal_spherical}.}

Ignoring the terms proportional to $\partial
\mathcal{F}^3_n(k)/\partial S_n$ in \ec{scaledepbiasgen} (which
vanishes for the local ansatz) we see that the coefficient of the $k^{-2}$ term, $c_n$, has a piece equal to the barrier times the Gaussian bias $(B_n b^{\rm g})$. This is the standard coefficient. However, the full treatment here has uncovered other terms, the remaining ones on the right in \ec{coeff}. In Fig. \ref{fig:bias_ellipsoidal_spherical} we plot $c_n$ for the spherical collapse barrier (i.e. $B(S_n) = \delta_c$) and for the ellipsoidal collapse barrier of \cite{ShethMoTormen:2001}:
\begin{align}\eql{ellcollapsebarrier}
B(S_n) = \sqrt{a} \delta_c \left[1 + \beta(a \nu)^{-\alpha} \right]
\end{align}
where $\nu = \delta_c^2 /S_n$, $a\approx0.7$, $\alpha\approx0.6$ and $\beta\approx0.4$.  For the spherical collapse barrier, $c_n$ is given by the well known expression $c_n=\delta_c^2/S_n - 1$ (e.g. \ec{biasspherical}).  For the ellipsoidal collapse model, we must use the more complicated expression we have derived in \ec{coeff}.

The naive prediction for $c_n$ (which holds in the case of the
spherical collapse barrier) is that it is related to the scale
independent bias via $c_n = 
b^{\rm g} B_n(S_n)$.  In fact, one way to try to extract constraints on $f_{\rm NL}$ is to fit the power spectrum with a bias of the form $1+b^{\rm g} + 2f_{\rm NL} b^{\rm g} \delta_c \mathcal{M}^{-1}(k)$. We have shown, however, in
Eq. \ref{eqn:scaledepbiasgen} that this simple relation does not appear to be correct
when the barrier evolves with the smoothing scale.  In
Fig. \ref{fig:bias_sn_dependence} we show the ratio of $c_n$
to the naive prediction as a function of $S_n$ for the ellipsoidal
collapse barrier given in \ec{ellcollapsebarrier}.    Note that our result depends heavily on the approximation of evaluating the 3-point functions at their end-point values, as in for example
  \ec{correlatorapprx}. This is known to break down as $S_n$ becomes large, of order $B_n^2$. Therefore, the upturn in Fig.~\rf{bias_sn_dependence}
  may be an artifact of our approximation and almost certainly does not persist as $S_n$ increases. Techniques that move beyond this approximation \cite{D'Aloisio:2012hr} are clearly called for.

\section{Conclusions}\label{sec:conclusions}

Making use of the path integral approach to excursion set theory, we have derived the bias of collapsed objects. We have kept all terms linear in the long wavelength density fluctuation and we have allowed for a completely general moving barrier. The bias contains both the scale independent contributions, analogous to those first reported by \cite{DeSimone:2011dn} for a general barrier, and contributions that depend on the long wavelength smoothing scale. We have  demonstrated how this dependence can be interpreted as a scale dependent biasing and shown that our result reduces to the famous Dalal et.\ al.\ \cite{Dalal:2007cu} result in the limit of spherical collapse and the local ansatz. In the spherical collapse limit for a general bispectrum shape, we reproduce the results of Desjacques, Jeong and Schmidt \cite{Desjacques:2011mq, Desjacques:2011jb}, including the previously overlooked additional term.

In the case where the barrier is not constant and depends on the smoothing scale, we find that the coefficient of the scale dependent bias is no longer simply related to the Gaussian bias parameter, but rather contains additional terms (\ec{coeff}) that might affect the extraction of $f_{\rm NL}$ from upcoming surveys. While the simple relation is recovered on sufficiently large mass scales, on smaller mass scales, we find a significant departure from the expected result. 

In arriving at this result, we have made several approximations. Following standard techniques in evaluating the effect of a moving barrier (which characterizes ellipsoidal collapse), we truncated the probability distribution in \ec{Piuncond}. This approximation causes $\Pi$ to no longer satisfy the Fokker-Planck equation. However, we have shown in Appendix \ref{app:linearbarrier} that when used in combination with the Fokker-Planck equation to calculate the halo bias, this approximation leads to results consistent with an exact treatment (at least to cubic order in derivatives) when compared to the case of a linear barrier where the probability distribution can be computed exactly. For this reason, we do not think this approximation leads to the upturn in $c_n$ at large $S_n$ depicted in Fig.\ \ref{fig:bias_sn_dependence}. 

A second approximation treating non-Gaussianity in the excursion set is likely to have more effect on our final answer.
We evaluated the 3-point function at the end point of the trajectories, as in \ec{correlatorapprx}. As noted there, one may think of this as keeping the zeroth order term in a Taylor expansion about the final time $S_n$. It is straightforward to calculate the contributions from additional terms in this series by making use of the results and methods of \cite{Maggiore:2009rv,Maggiore:2009rw,Maggiore:2009rx}. This approximation can be shown to correspond to an expansion in $S_n/B_{n}^2$. Furthermore, using the saddle point techniques of \cite{D'Amico:2010ta} it is possible to resum a number of higher order corrections to this formula to obtain a more accurate formula for the bias. However, to reproduce the existing results in the literature, the zeroth order approximation used here appears sufficient. We leave the calculation of higher order corrections to future work.

While this paper was in preparation, we became aware of the work \cite{D'Aloisio:2012hr}, which also considers the halo bias in the path integral formulation of the excursion set. While the authors of \cite{D'Aloisio:2012hr} consider only a constant barrier, they include next-to-leading order corrections from relaxing the approximation at \ec{correlatorapprx}. The results of this work are consistent with those found  in \cite{D'Aloisio:2012hr} the limit where the barrier is constant, and one restricts to the leading order result.

\acknowledgements
We thank Neal Dalal, Anson D'Aloisio, Wayne Hu, Donghui Jeong, and Toni Riotto for useful discussions and the authors of \cite{D'Aloisio:2012hr} for sharing an earlier draft of their work with us. This work was supported in part by National Science
Foundation under Grant AST-090872, the Kavli Institute for Cosmological Physics at the University of Chicago through grants NSF PHY-0114422 and NSF PHY-0551142 and an endowment from the Kavli Foundation and its founder Fred Kavli.  
SD is supported by the U.S.
Department of Energy, including grant DE-FG02-95ER40896. AL was supported in part by the NSF through grant
AST-1109156.

\begin{appendix}

\section{The linear barrier}\label{app:linearbarrier}

In this Appendix we compare our results to a case where the probability distribution can be evaluated exactly. Aside from the limit where the absorbing barrier is independent of the smoothing scale, the only other known exact solution to the Fokker-Planck equation with an absorbing barrier condition is in the case where the barrier has a linear dependence on the smoothing scale,
\begin{align}\eql{linbar}
B(S) = B_0 + B_1 S
\end{align}
where $B_0$ and $B_1$ are constants. For this barrier, it is an elementary exercise to obtain the probability distribution 
\begin{align}\nonumber
\Pi^{\rm LB}(\delta_n, S_n)  = & \frac{1}{\sqrt{2\pi S_n}}\bigg(\exp\left[-\frac{\delta_{n}^2}{2S_n}\right]\\ \nonumber& -\exp\left[-\frac{(2B_n - \delta_n)^2}{2S_n}+2B_{n}'\left(B_n - \delta_{n} \right) \right] \bigg),
\end{align}
which one can verify solves the Fokker-Planck equation, with the linear boundary condition \ec{linbar}. The superscript $^{\rm LB}$ here and in the rest of this appendix denotes expressions that are only valid for the barrier in \ec{linbar}. 

It is straightforward to obtain the Gaussian formation rate,
\begin{align}
\mathcal{F}^{\rm LB}(S_n) = & \frac{B_n - S_n B'_n}{\sqrt{2\pi S^3}}e^{-\frac{B_n^2}{2S_n}}
\end{align}
which agrees exactly with the result in \ec{fpcross} since for the barrier in \ec{linbar}
\begin{align}
\mathcal{P}_{0n} = -S_n B'_n.
\end{align}
The conditional crossing rate is found as described above by simply translating the Gaussian result using the shift of variable in \ec{shift}. We can also compute the halo bias analogous to \ec{diasrealspace}, 
\begin{widetext}
\begin{align}\nonumber
\delta^{\rm halo, LB}_{m} = &\(\frac{B_n}{S_n}-\frac{1}{(B_n- S_n B_n')}\)\delta_m\\\nonumber
& - \frac{1}{6}\frac{\mathcal{S}_{n^3}}{(B_n - S_n B_n')^2}\Bigg(-20S_n B_n'B_n+10 B_n^2-9(S_n B_n')^2-3 \frac{B_n^2}{S_n} \left(B_n- S_n B_n'\right){}^2+7 S_n\\\nonumber
& -8 \sqrt{2 \pi } S_n^{5/2} B_n'{}^3 e^{\frac{\left(B_n-2 S_n B_n'\right){}^2}{2 S_n}} \left(\left(B_n-2
   S_n B_n'\right) \left(B_n-S_n B_n'\right)-S_n\right) \text{erfc}\left[\frac{B_n-2 S_n B_n'}{\sqrt{2S_n}}\right]
   \Bigg)\delta_m\\\nonumber
  &  -\frac{S_n\mathcal{S}'_{n^3}}{3(B_n- S_n B_n')^2}\Bigg( (B_n + S_n B_n')(B_n-3S_n B_n')+S_n- 4  S_n^2 B_n'^3 \left(B_n- S_n B_n'\right)\\ \nonumber
  &+2 \sqrt{2 \pi } S_n^{3/2} B_n'^3 e^{\frac{\left(B_n - 2S_n B_n'\right){}^2}{2 S_n}}
   \left(\left(B_n- S_n B_n'\right) \left(B_n- 2S_n B_n'\right)-S_n\right) \text{erfc}\left(\frac{B_n-2S_n B_n'}{\sqrt{2S_n}}\right)\Bigg)\delta_m\\ \nonumber
&\frac{\mathcal{S}_{n^2 m} }{2} \left(\frac{B_n^2}{S_n}-\frac{B_n}{(B_n- S_n B_n')}+\frac{3S_n B_n'}{(B_n- S_n B_n')}+2\sqrt{2\pi S_n^3}\frac{B_n'{}^2 e^{\frac{\left(B_n-2 S_n B_n'\right){}^2}{2 S_n}}}{(B_n- S_n B_n')} \text{erfc}\left[\frac{B_n-2 S_n B_n'}{\sqrt{2 S_n}}\right]\right)\frac{\delta_m}{\sqrt{S_m}}\\
& +S_n\mathcal{S}'_{n^2 m}\left( 1+\frac{2S_n B_n'}{(B_n- S_n B_n')} + \sqrt{2\pi S_n^3}\frac{B_n'{}^2 e^{\frac{\left(B_n-2 S_n B_n'\right){}^2}{2 S_n}}}{(B_n- S_n B_n')} 
   \text{erfc}\left[\frac{B_n-2 S_n B_n'}{\sqrt{2S_n}}\right]\right) \frac{\delta_m}{\sqrt{S_m}}
\end{align}
from which we can obtain the scale dependent part of the bias
\begin{align}\nonumber
b^{LB}(k) = & \frac{\mathcal{F}^{3}_{n}(k)}{2\mathcal{M}(k)}\left(B_n\(\frac{B_n}{S_n}-\frac{1}{B_n- S_n B_n'}\)+3\frac{S_n B_n'}{B_n- S_n B_n'} +2\sqrt{2\pi S_n^3}\frac{B_n'{}^2 e^{\frac{\left(B_n-2 S_n B_n'\right){}^2}{2 S_n}}}{B_n- S_n B_n'}\text{erfc}\left[\frac{B_n-2 S_n B_n'}{\sqrt{2S_n}}\right]\right)\\
& +\frac{1}{ \mathcal{M}(k)}\frac{d \mathcal{F}^{3}_{n}(k)}{d\ln S_n}\left(1+\frac{2S_n B_n'}{(B_n- S_n B_n')} +\sqrt{2\pi S_n^3}\frac{B_n'{}^2 e^{\frac{\left(B_n-2 S_n B_n'\right){}^2}{2 S_n}}}{B_n- S_n B_n'}
   \text{erfc}\left[\frac{B_n-2 S_n B_n'}{\sqrt{2S_n}}\right]\right)
\end{align}
\end{widetext}
and thus we find exact agreement with \ec{diasrealspace} and \ec{scaledepbiasgen} up to terms cubic in $B_n'$, the order to which we have evaluated the approximation in \ec{Piuncond}. We thus conclude that the deviation we find at low masses is not a consequence of the approximation of the probability distribution, but rather represents a break down of the approximations at \ec{correlatorapprx} or possibly of the excursion set itself.

\end{appendix}

\bibliography{ScaleDepBias}
\end{document}